\documentclass[prb,preprint]{revtex4-1} 

\usepackage{amsmath}  
\usepackage{amsfonts} 
\usepackage{graphicx} 
\usepackage[T1]{fontenc}
\usepackage{lmodern}
\usepackage{color}
\usepackage{hyperref}
\usepackage{epstopdf}
\usepackage[table]{xcolor}
\usepackage{bm}
\usepackage{braket}
\usepackage{matlab}
\usepackage[font = footnotesize]{caption}
\usepackage[font = scriptsize]{subcaption}
\usepackage{float}
\usepackage{booktabs}
\begin{document}
\setlength{\parindent}{1em}

\title{Path integral approach to driven quantum harmonic oscillator using Markov chain Monte Carlo methods}

\author{Sohini Marik}
\author{Souvik Naskar}
\thanks{The authors Sohini Marik and Souvik Naskar contributed equally.}
\author{Shibaji Banerjee}
\affiliation{Department of Physics, St. Xavier's College, Kolkata}



\date{\today}

\begin{abstract}
We have simulated the ground states of quantum harmonic oscillators driven either by constant forces of different magnitudes or time-dependent driving forces. The expectation values of position for various combinations of mass, natural angular frequency, and the coupling constant $\lambda$ were calculated for both driving modes. For constant forcing, coherent states were obtained. The results for both forcing scenarios match the theoretically expected values almost exactly. For the simulations, the Metropolis algorithm was implemented on a discrete time lattice to evaluate the imaginary time path integral of the systems. 
\end{abstract}


\maketitle
\section{Introduction}
In the Schr\"{o}dinger formulation of quantum mechanics developed in 1925, the time evolution of a non-relativistic system is controlled by its Hamiltonian. The path integral formulation is an alternate approach that relies on a system's Lagrangian as the fundamental quantity. It is the  generalization of the classical action principle to quantum mechanics. Most of this formulation was developed by R.P. Feynman in 1948. A precursor to his work was P. Dirac's 1933 paper that proposed an analogy between the complex exponential of the Lagrangian and the transformation function relating quantum mechanical wave functions at consecutive instants of time. Feynman calculated that the complex exponential of the action integrated over all possible trajectories of a particle between two space-time points $\left( x_i, t_i \right) $ and $ \left( x_f, t_f \right) $  yields the probability amplitude that the particle at $x_i$ at $t_i$ will be at $x_f$ at $t_f$\cite{feynman,feynmanhibbs,blundell}.

One of the few exactly solvable path integrals is that of the driven harmonic oscillator system. \cite{ingold,jana} 
Driven harmonic oscillators have been extensively used in literature to approach various problems in physics. For instance, Piilo and Maniscalco simulated a non-Markovian damped oscillator with a forced harmonic oscillator in 2006\cite{opensys}, and Gimelshein \textit{et al} applied a 3D forced harmonic oscillator model of vibration-translation energy to atomic and molecular collisions in 2017\cite{vib}.
The expression of the probability amplitude of a one-dimensional harmonic oscillator driven by a general time-dependent force is provided in Feynman and Hibbs\cite{feynman}. In this paper, we will take a non-perturbative computational approach to study the ground state probability distribution of the system when the driving force is beyond the perturbative limit. We have specifically considered two cases --- a constant force and a sinusoidal force. We have simulated the ground states of both systems using Markov Chain Monte Carlo methods to evaluate the imaginary time path integral. A similar procedure has been implemented by Westrbroek \textit{et al}\cite{westbroek} and Mittal \textit{et al}\cite{mittal} to compute the ground state of a simple harmonic oscillator and an anharmonic oscillator respectively. We have further demonstrated that the ground state of the driven harmonic oscillator can be described by coherent states, and compared our simulations with the theoretical result of the position expectation value obtained from Carruthers and Nieto's 1965 paper\cite{carruthers}.

The organization of our paper is as follows. We have briefly introduced the Feynman path integrals, coherent states, and the driven quantum harmonic oscillator in the context of our paper in sections \ref{section:pi}, \ref{section:cs}, and \ref{section:fho} respectively. In section \ref{section:mcmc}, we have implemented Markov chain Monte Carlo (MCMC) methods in two driven harmonic oscillator systems and analyzed the results.
\section{Feynman Path Integrals}
\label{section:pi}
\begin{table}[ht]
    \centering
    \begin{tabular}{l  c}\hline
       \bf{Parameter}  & \bf{Meaning} \\
       \hline
        $\tau$ & Imaginary time $\tau = it$\\
        $\delta \tau$ & Lattice spacing in discrete imaginary time lattice\\
        $N_{\tau}$ & Length of discrete imaginary time lattice\\
        $\lambda$ & Coupling constant for driving force in driven harmonic oscillator\\
        $\tilde{m}$ & Dimensionless mass $\tilde{m} = m\delta \tau$\\
        $\tilde{\omega}$ & Dimensionless frequency $ \tilde{\omega} = \omega \delta \tau$\\
        $\tilde{x}_i$ & Dimensionless position on discrete time lattice $\tilde{x}_i = \dfrac{x_i}{\delta \tau}$\\
        $\tilde{F}_i$ & Dimensionless driving force on discrete time lattice $\tilde{F}_i = F_i \left(\delta\tau\right)^2$\\
        $\tilde{S}$ & Dimensionless Euclidean action\\
        $\alpha$ & Expectation value of position of the coherent state $\ket{\alpha}$\\
        \hline
    \end{tabular}
    \caption{An overview of the notation used.}
    \label{tab:notn}
\end{table}
In the Lagrangian formulation of classical mechanics, the trajectory of a particle is given by the solutions of the Euler-Lagrange equations. This path minimizes the classical action $S_{cl}=\int Ldt,$ where $L$ is the Lagrangian of the system. In quantum mechanics, the particle is not restricted to a single trajectory. It can go from one point to another by all accessible paths. Each path contributes a phase related to the classical action. To compute the probability amplitude, we have to sum over all these phase factors. The propagator of a particle of mass $m$ going from $x_i$ at time $t_i$ to $x_f$ at time $t_f$ in a potential $V(x(t))$ is given by
\begin{equation}
    \langle x_f ,t_f |x_i ,t_i \rangle =\int Dx(t)e^{iS/\hbar },
\end{equation}
where the action of the path $x(t)$ is
\begin{equation}
    S=\int_{t_i }^{t_f }{ dt\left\lbrack \frac{1}{2}m{\left(\frac{dx}{dt}\right)}^2 -V\left(x(t)\right)\right\rbrack}.
\end{equation}
In the limit $t \rightarrow -i\tau$ ($\tau$ is a real number), we get the Euclidean time integral
\begin{equation}
     \langle x_f ,t_f |x_i ,t_i \rangle =\int Dx(t)e^{-S_E/\hbar },
\end{equation}
where
\begin{equation}
    S_E=\int_{\tau_i }^{\tau_f } d\tau \left\lbrack \frac{1}{2}m{\left(\frac{dx}{d\tau }\right)}^2 +V\left(x(\tau )\right)\right\rbrack
\end{equation}
is the Euclidean action. This form of the integral is not oscillatory. Also, it is damped and the contributions of the higher energy states become negligible for large values of $\tau$.\cite{blundell} In this paper, we will apply Monte Carlo Markov Chain methods to the Euclidean time integral and compute the ground state of a driven harmonic oscillator.

\section{Coherent states}
\label{section:cs}
Coherent states are the states of a quantum harmonic oscillator that show classical behavior.\cite{blundell} A coherent state $\ket{\alpha}$ is defined as
\begin{equation}\label{coherent}
    \ket{\alpha} \equiv T_{\alpha}\ket{0} = \exp\left(-\frac{i}{\hbar}\hat{p}\alpha\right) \ket{0},
\end{equation}
where $\hat{p}$ is the momentum operator, and $\ket{0}$ is the ground state of the simple harmonic oscillator. The multiplication property of the translation operator $T_{\alpha}$ is the following:
\begin{equation}
    T_{\alpha}T_{\beta} = \exp\left(-\frac{i}{\hbar}\hat{p}\alpha\right)\exp\left(-\frac{i}{\hbar}\hat{p}\beta\right) = \exp\left(-\frac{i}{\hbar}\hat{p}\left(\alpha + \beta\right) \right) = T_{\alpha + \beta}.
\end{equation}
It follows from the above multiplication property that
\begin{equation}
     T_{\alpha}^{\dagger} = \exp\left(\frac{i}{\hbar}\hat{p}\alpha\right) = \exp\left(-\frac{i}{\hbar}\hat{p}(-\alpha)\right) = T_{-\alpha}= T_{\alpha}^{-1},
\end{equation}
establishing $T_{\alpha}$ is unitary. It is called the translation operator due to its action on the position operator $\hat{x}$:
\begin{equation}
    T_{\alpha}^{\dagger}\hat{x}T_{\alpha} = \exp\left(\frac{i}{\hbar}\hat{p}\alpha\right)\hat{x}\exp\left(-\frac{i}{\hbar}\hat{p}\alpha\right)= \hat{x}+\frac{i}{\hbar}\left[\hat{p}, \hat{x}\right]\alpha = \hat{x} + \alpha.
\end{equation}
Consequently, the position expectation value of the coherent state $\ket{\alpha} $ is 
\begin{equation}
    \bra{\alpha}\hat{x}\ket{\alpha} = \bra{0}T_{\alpha}^{\dagger}\hat{x}T_{\alpha}\ket{0} = \bra{0}\hat{x} + \alpha \ket{0} = \alpha.\cite{zwiebach}
\end{equation}

\section{Driven Harmonic Oscillator}
\label{section:fho}
The expression of the action of a driven harmonic oscillator as a function of the path $x(t)$ is 
\begin{equation}
    S=\int dt\left\lbrack \frac{1}{2}m\dot{x} (t)^2 -\frac{1}{2}kx(t)^2 + x(t)F(t)\right\rbrack\cite{blundell}.
\end{equation} 
We can introduce a coupling constant $\lambda$ to scale the forcing term. Then the action becomes
\begin{equation}
    S=\int dt\left\lbrack \frac{1}{2}m\dot{x} (t)^2 -\frac{1}{2}kx(t)^2 + \lambda x(t)F(t)\right\rbrack. \label{eq:action}
\end{equation}

According to Carruthers and Nieto\cite{carruthers}, the ground state $\ket{0}'$ of the driven harmonic oscillator with a constant driving force $F_0$ is 
\begin{equation}\label{newground}
    \ket{0}' = \exp\left[\frac{x_0F_0}{\hbar \omega} \left(\hat{a}^{\dagger} - \hat{a}\right)\right]\ket{0},
\end{equation}
where $x_0 = \sqrt{\dfrac{\hbar}{2m\omega}}$, $\hat{a}$ is the annihilation operator, and $\hat{a}^{\dagger}$ is the creation operator. Now, using the definition of $\hat{p}$:
\begin{equation}
    \hat{p} = -im\omega x_0 \left(\hat{a} - \hat{a}^{\dagger} \right),
\end{equation}
and substituting that into equation \eqref{newground}, we get
\begin{equation}\label{eq:cohground}
    \ket{0}' = exp\left[-\frac{i}{\hbar}\hat{p}\left\{\frac{F_0}{m\omega^2}\right\}\right]\ket{0}
\end{equation}.
Now comparing \eqref{coherent} with \eqref{eq:cohground}, we can see that the ground state $\ket{0}'$ of the driven harmonic oscillator is a coherent state with the expectation value of position $\alpha$ given by 
\begin{equation}
    \alpha = \frac{F_0}{m\omega^2}. \label{eq:alpha}
\end{equation}

\begin{figure}[ht]
    \centering
    \includegraphics[width = 8cm]{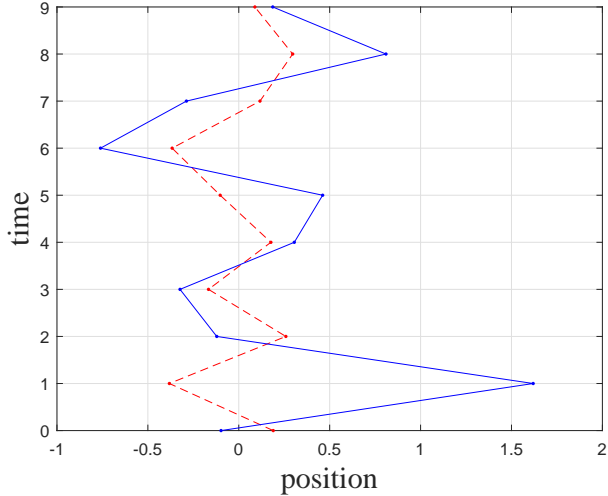}
    \caption{Update of a discrete imaginary-time path between $\tau_i = 0$ to $\tau_i = 9$, illustrating the computational method. The red path represents the thermalized path $\text{path}^{(0)}$ and the path in blue is one after 5 sweeps. }
    \label{fig:update}
\end{figure}
\section{Implementation of Monte Carlo methods}\label{section:mcmc}
The Metropolis algorithm has been implemented on a discrete-time lattice with $N_{\tau}$ time slices and a periodic boundary condition $N_{\tau}+1=1$. The Euclidean time is $\tau_i = i \delta \tau $ where $i \in \lbrace 1, \dots N_{\tau} \rbrace$ is the site index, and $\delta \tau$ is the lattice spacing. 
The trajectory over the time lattice is characterized by a real number array $(x_1 ,\dots,x_N )$. First, we want to express all relevant physical quantities as real numbers. For this, we set $ \hbar =1=c$. It follows that $[\textrm{time}]=[\textrm{length}]=[{\textrm{mass}}^{-1} ]=[{\textrm{energy}}^{-1} ]$. Introducing the dimensionless variables
\begin{equation}
    \tilde{m} =m\delta \tau ,~\tilde{\omega} =\omega \delta \tau ,~{\tilde{x} }_i =\frac{x_i }{\delta \tau },~{\tilde{F} }_i =F_i {\left(\delta \tau \right)}^2
\end{equation}
in terms of the lattice spacing $\delta \tau$, we get the discrete form of the dimensionless action as 
\begin{equation}
    \displaystyle{ \tilde{S} =\sum_{i=1}^N \frac{1}{2}\tilde{m} {\left({\tilde{x} }_{i+1} -{\tilde{x} }_i \right)}^2 +\frac{1}{2}\tilde{m} {\tilde{\omega} }^2 {\tilde{x} }_i^2 -{\tilde{x} }_i {\tilde{F} }_i }.
\end{equation}
The initial configuration ${\textrm{path}}^{(0)}$ is updated by the Metropolis algorithm to get the next configuration ${\textrm{path}}^{(1)}$, and so on. One update to the value of the path $x_i$ at the lattice site $i$ constitutes one Monte-Carlo step. The lattice sites are randomly visited. A new value $x_i^{(\textrm{new})} =x_i +u$ is proposed from a symmetric normal distribution around the previous value $x_i$. As $x_i^{(\textrm{new})}  $ always depends on $x_i$, the propositions are autocorrelated.  There are $N$ Monte-Carlo steps in one Metropolis sweep. In each sweep, every lattice site gets updated once on average. To reduce autocorrelation, we have discarded a number of sweeps between every two path configurations that would be utilized for simulation.\cite{westbroek}

In the random walk Metropolis algorithm, the probability $P$ of proposing a state $b$ from the current state $a$ is 
\begin{equation}
    P=\frac{\pi (b)}{\pi (a)},
\end{equation}
where $\pi$ is the target probability distribution. The probability of accepting the proposition is $\min \left\lbrace P,\,1\right\rbrace$. The candidate states proposed in the initial metropolis sweeps are not from the target distribution. The number of sweeps required to reach the target distribution is called the burn-in period. Once the target distribution is reached, the mean and standard deviation of the paths proposed in each sweep becomes nearly constant.\cite{guilhoto,hastings,creutz} In our case, $\pi$ is $e^{-S_E}$, and the acceptance rate is $\min \left\lbrace e^{-\delta S_E},1\right\rbrace$. $\delta S_E$ is the change in action due to the proposed change in path. Thus, we always accept propositions that decrease the action. Propositions that increase the action are accepted with a probability of $e^{-\delta S_E}$. The function used to simulate each metropolis sweep is provided in the appendix. We have used an array of random numbers as the initial configuration. This is called a hot start. We could also use a cold start with
an array of zeros. The choice of initial configuration doesn't make a difference after the burn-in period. 
Based on Westbroek \textit{et al}\cite{westbroek}, a time lattice with 120 lattice points has been chosen. We have taken a grand total of 12,000 sweeps discarding 12 sweeps in between every accepted configuration after the burn-in period. The lattice can be made finer by increasing the number of lattice points, but that significantly increases computation time.
\begin{figure}[ht]
     \centering
     \begin{subfigure}[b]{0.5\textwidth}
         \centering
         \includegraphics[width=7cm]{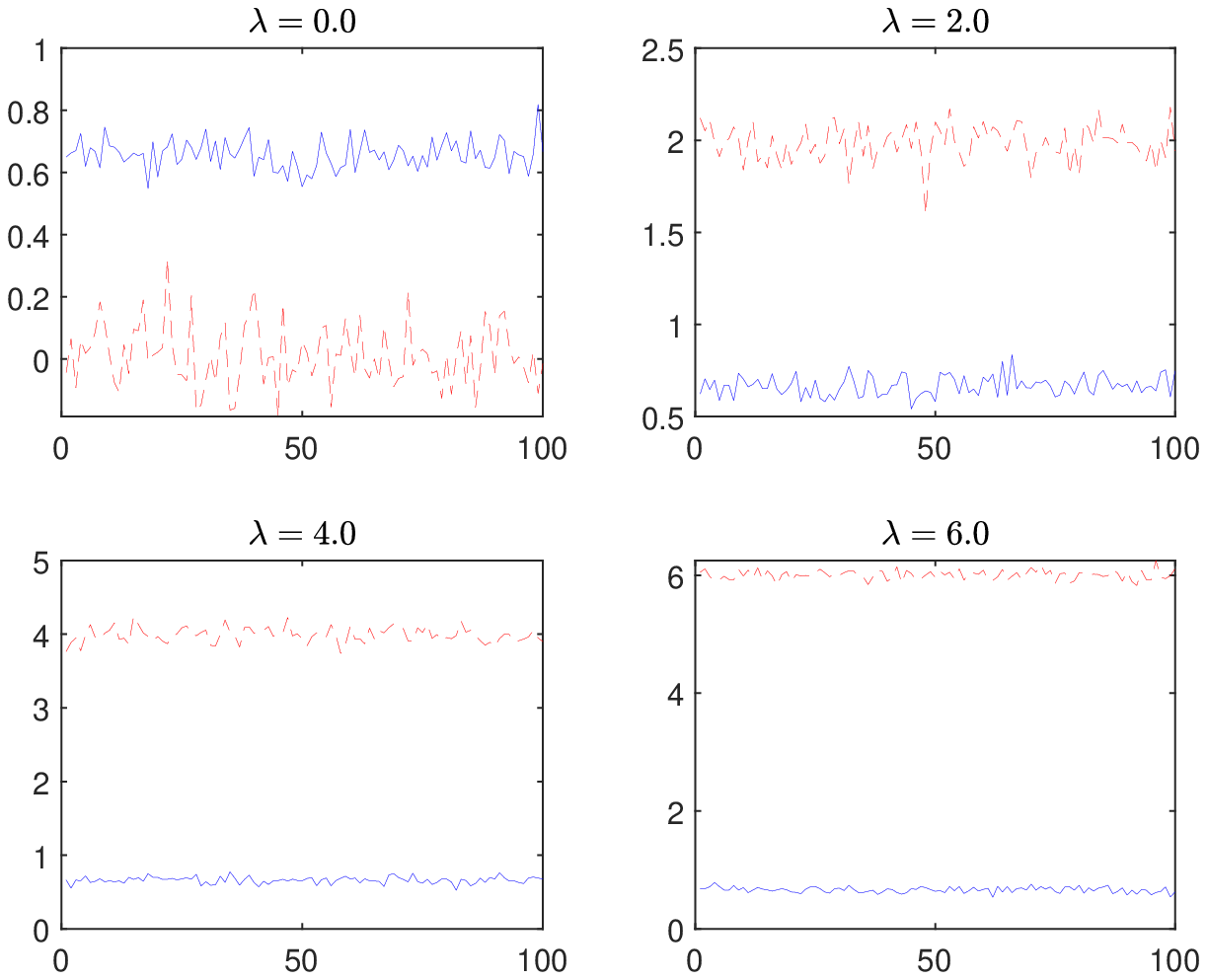}
         \subcaption{}
         \label{fig:cburn}
     \end{subfigure}%
     \hfill
     \begin{subfigure}[b]{0.5\textwidth}
         \centering
         \includegraphics[width=7cm]{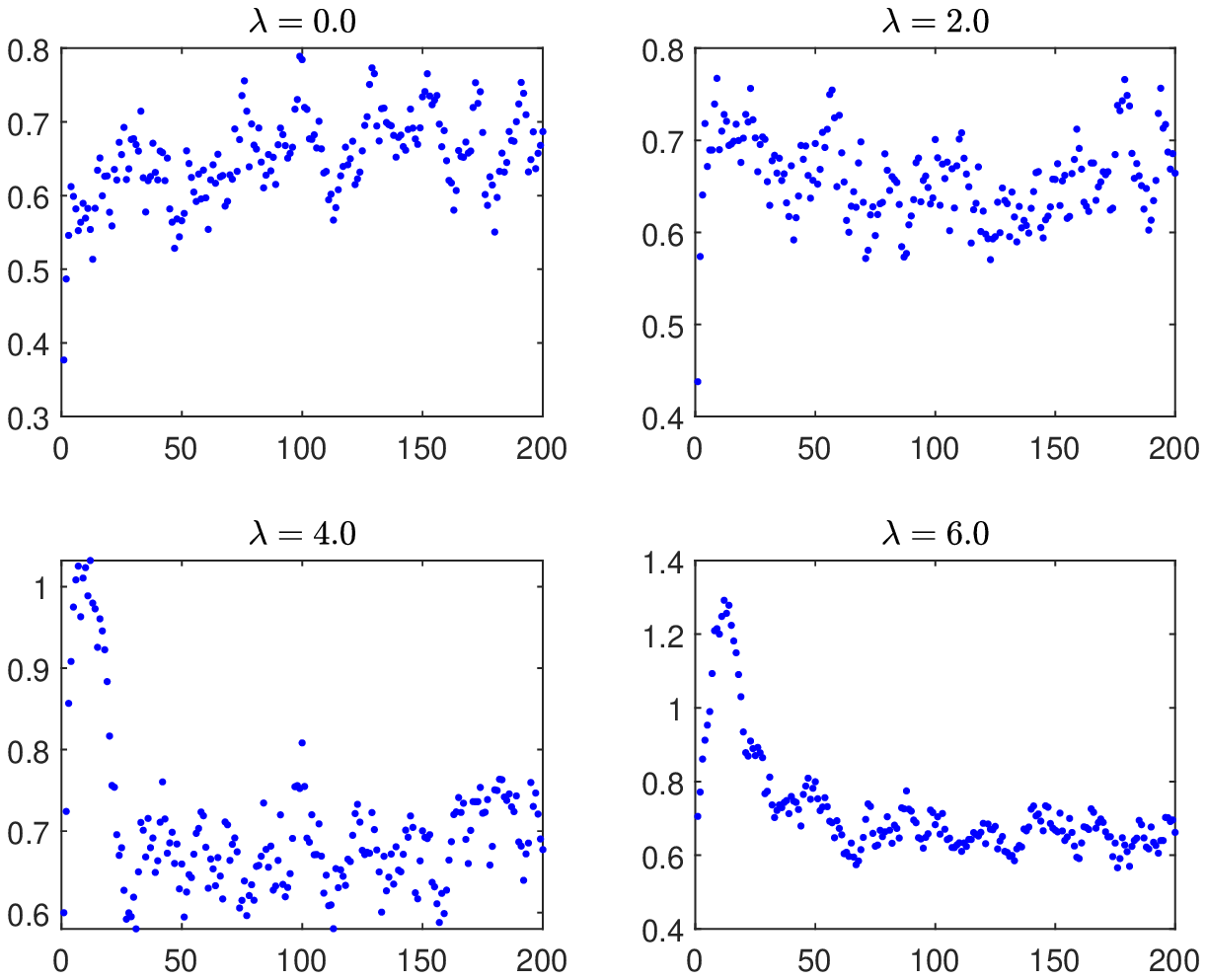}
         \caption{}
         \label{fig:cburnz}
     \end{subfigure}%
        \caption{\ref{fig:cburn} shows a trial run for the mean $\langle x \rangle$ (dashed red line) and the standard deviation $\sqrt{\langle x^2 \rangle - {\langle x \rangle}^2}$ (solid blue line) to illustrate the thermalization effects $\left(\tilde{m} = 1,\ \tilde{\omega} = 1,\ N_{\tau} = 120\right)$. 120 paths were discarded between every shown path. \ref{fig:cburnz} shows the standard deviation for the same trial run for the first 200 sweeps out of the 12,000 to illustrate the burn-in period. In all the cases the algorithm reaches the target distribution before 100 sweeps. }
        \label{fig:cburnin}
\end{figure}

\subsection{Constant Driving Force} \label{section:mcmcfho}
In this section, we will consider a constant driving force for different values of the coupling constant $\lambda$. In the discrete form, $\tilde{F}_i = \lambda, \ \forall \ i, \ i \in \lbrack 1,120 \rbrack$. The action in equation \eqref{eq:action} becomes,
\begin{equation}\label{eq:consact}
    \displaystyle{ \tilde{S} =\sum_{i=1}^N \frac{1}{2} {\tilde{m}\left({\tilde{x} }_{i+1} -{\tilde{x} }_i \right)}^2 +\frac{1}{2} \tilde{m}\tilde{\omega}^2{\tilde{x} }_i^2 -{\tilde{x} }_i {\lambda } }.
\end{equation}
We have assessed the burn-in period from a trial simulation by plotting the standard deviations of the proposed paths vs the number of sweeps taking $\tilde{m} =1$ and $\tilde{\omega} = 1$. From figure \ref{fig:cburnin}, we see that the target distribution is reached in the first 50-100 sweeps. So, we have taken the burn-in period as 100 sweeps for all cases. 
\begin{figure}[ht]
     \centering
     \begin{subfigure}[b]{0.5\textwidth}
         \centering
         \includegraphics[height=6cm]{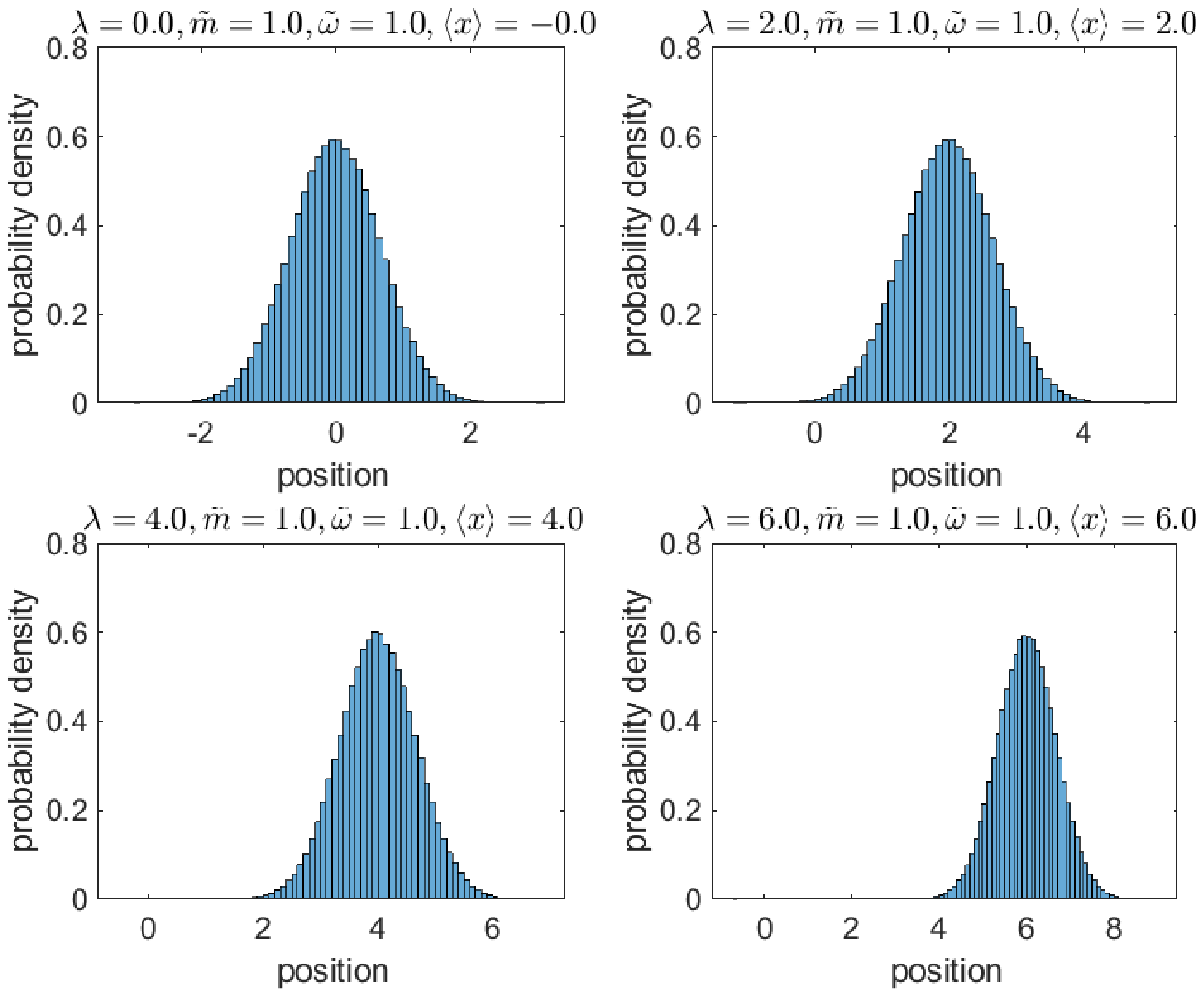}
         \caption{}
         \label{fig:cgs}
     \end{subfigure}%
     \hfill
     \begin{subfigure}[b]{0.5\textwidth}
         \centering
         \includegraphics[height=6cm]{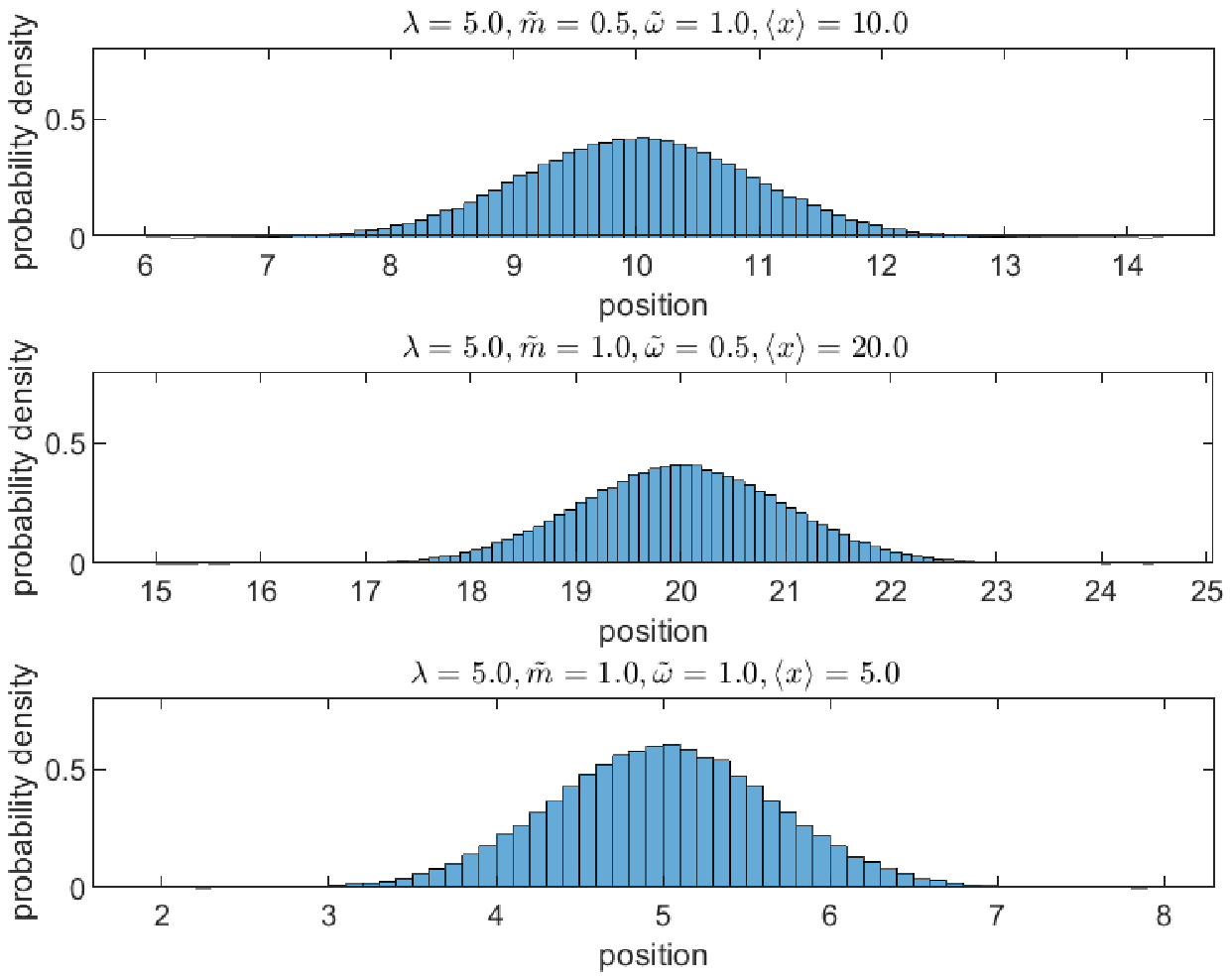}
         \caption{}
         \label{fig:cgsmw}
     \end{subfigure}%
        \caption{\ref{fig:cgs} shows the ground state probability distribution for different values of the coupling constant $\lambda$ keeping the values of the dimensionless mass $\tilde{m}$ and dimensionless frequency $\tilde{\omega}$ both at $1$. \ref{fig:cgsmw} shows the ground state for different combinations of $\tilde{m}$ and $\tilde{\omega}$ keeping $\lambda$ fixed at $5$. In all these cases we see that $\langle x \rangle$ takes the value $\alpha$ of equation \eqref{eq:calpha}. All simulations were done with 120 time lattice points and 12,000 metropolis sweeps. 12 paths were discarded between every two accepted paths.}
        \label{fig:cgsstuff}
\end{figure}

For the ground state simulation, we have plotted the histogram of all the paths generated by the algorithm, barring the burn-in data. Figure \ref{fig:cgs} shows the normalized ground states $\left| \psi_0 \right|^2$ for $\tilde{m}=1$, $\tilde{\omega} = 1$, and $\lambda = 0, 2, 4, 6$. For $\lambda = 0$, we have retrieved the ground state of a simple harmonic oscillator obtained by Westbroek \textit{et al}. The ground states associated with the other values of $\lambda$ have the same waveform, but they are displaced towards the right. This illustrates that these ground states are coherent states. The shift can be calculated from the position expectation value of a coherent state given in equation \eqref{eq:alpha}. Substituting the expression of the forcing function in equation \eqref{eq:alpha}, we get
\begin{equation}
    \alpha = \dfrac{\lambda}{\tilde{m} \tilde{\omega }^2}. \label{eq:calpha}
\end{equation}
 In figure \ref{fig:cgsmw}, we have plotted the ground state probability distribution for different combinations of $\tilde{m}$ and $\tilde{\omega}$, keeping $\lambda$ fixed at $\lambda=5$. The values of $\alpha $ predicted by equation \eqref{eq:calpha} are --- $\alpha = 5$ for $\tilde{m} = 1$ and $\tilde{\omega} = 1$, $\alpha = 10$ for $\tilde{m} = 0.5$ and $\tilde{\omega}= 1$, and $\alpha = 20$ for $\tilde{m} = 1$ and $\tilde{\omega} = 0.5$. These are precisely the values of $\alpha$ we have got from our simulations in figure \ref{fig:cgsmw}.
\begin{table}[ht]
    \centering
    \begin{tabular}{|c|c|c|c|}
    \hline
      $\bm{\lambda}$   & $\bm{\tilde{m}}$ & $\bm{\tilde{\omega}}$ & $\bm{\langle x \rangle}$ \\
      \hline
       0.0  & 1.0  &  1.0  &  -0.0 \\
       \hline
       2.0 & 1.0 & 1.0 & 2.0\\
       \hline
       4.0 & 1.0 & 1.0 & 4.0\\
       \hline
       6.0 & 1.0 & 1.0 & 6.0\\
       \hline
       5.0 & 1.0 & 1.0 & 5.0\\
       \hline
       5.0 & 0.5 & 1.0 & 10.0\\
       \hline
       5.0 & 1.0 & 0.5 & 20.0\\
       \hline
    \end{tabular}
    \caption{This is a table of values of the mean position $\langle x \rangle$ for various combinations of the coupling constant $\lambda$, dimensionless mass $\tilde{m}$ and dimensionless frequency $\tilde{\omega}$. At every instance, $\langle x \rangle$ takes the value of $\alpha$ from equation \eqref{eq:calpha}. Equation \eqref{eq:calpha} describes the expectation value of position of a coherent state. This establishes that the ground state of a driven harmonic oscillator with a constant driving force is a coherent state. }
    \label{tab:meanx}
\end{table}
\subsection{Sinusoidal Driving Force}
In this section, we will study a harmonic oscillator potential driven by a sinusoidal forcing function. The general form of the force is
\begin{equation}
    \tilde{F}_i = \sin \left( \tilde{\omega}_d \tau_i - \tilde{\phi} \right). \label{eq:form}
\end{equation}
The choice of $\tilde{\omega}_d$ should ensure that $F_{N+1} = F_1$, satisfying the periodic boundary condition imposed on the time lattice. In our simulation, we have taken $ \tilde{\omega}_d = \dfrac{\pi \tilde{\omega}}{12} $, and $\tilde{\phi} = 0$. Substituting these values in equation \eqref{eq:form}, the action in equation \eqref{eq:action} becomes,
\begin{equation}
     \tilde{S} =\sum_{i=1}^N \frac{1}{2}\tilde{m} {\left({\tilde{x} }_{i+1} -{\tilde{x} }_i \right)}^2 +\frac{1}{2}\tilde{m} {\tilde{\omega} }^2 {\tilde{x} }_i^2 -{\tilde{x} }_i  \lambda \sin \left(\dfrac{\pi}{12}  \tilde{\omega} \tau_i \right).
\end{equation}

Again, the burn-in period is 100 sweeps. In this case, the histogram of all simulated path configurations will give us the ground state probability distribution averaged over time. Figure \ref{fig:fgs} shows the ground states for $\lambda = 0, 2, 4, 6$. For $\lambda>0$, we get two maxima in the distribution. As $\lambda$ increases, the maxima shift away from zero symmetrically. 
\begin{figure}[ht]
    \centering
    \includegraphics[width=10cm]{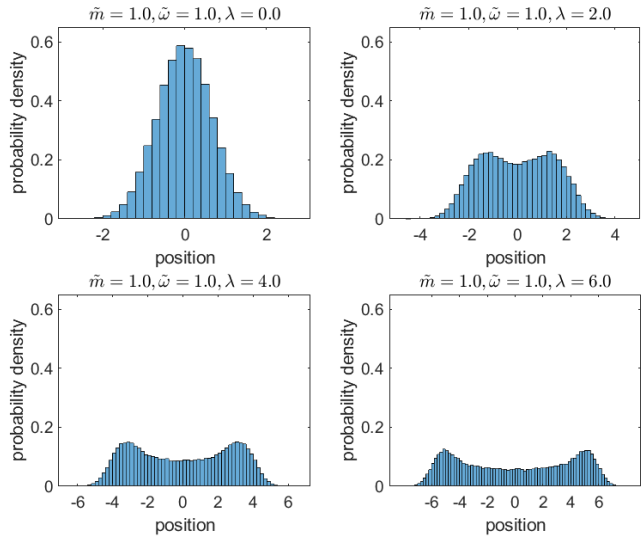} \label{fig:fgs}
    \caption{The figure shows the ground state probability distribution of a driven harmonic oscillator with a sinusoidal driving force for different values of the coupling constant $\lambda$. As $\lambda$ is turned on and increased, split peaks occur near the values $+\alpha$ and $-\alpha$. Here, $\alpha$ is the position expectation value calculated from equation \eqref{eq:calpha}. All simulations were done with 120 time lattice points and 12,000 metropolis sweeps. 12 paths were discarded between every two accepted paths. }
    \label{fig:fgs}
\end{figure}

The driving force has an explicit time dependence. This implies that $F_i$ has a constant value at each lattice point $\tau_i$ in the discrete form. Therefore, the ground state probability density is a coherent state for every $\tau_i$. As in the previous case, we can calculate the displacement of the coherent states from equation \eqref{eq:alpha}. Using $\tilde{F}_i =  \lambda \sin \left( \dfrac{\pi}{12}
 \tilde{\omega} \tau_i \right)$ in equation \eqref{eq:alpha}, we get
\begin{equation}
    \alpha_i = \frac{\lambda}{\tilde{m} \tilde{\omega}^2 } \sin \left(\frac{\pi\tilde{ \omega}}{12}  \tau_i \right). \label{eq:fal}
\end{equation}
So, the displacement of the coherent states as a function of $\tau_i$ is a sinusoidal curve with the same frequency and initial phase as that of the driving force. The amplitude of the curve is the maximum displacement of the mean position of the coherent state from $\langle x \rangle = 0 $. 

\begin{figure}[ht]
    \centering
    \includegraphics[width=10cm]{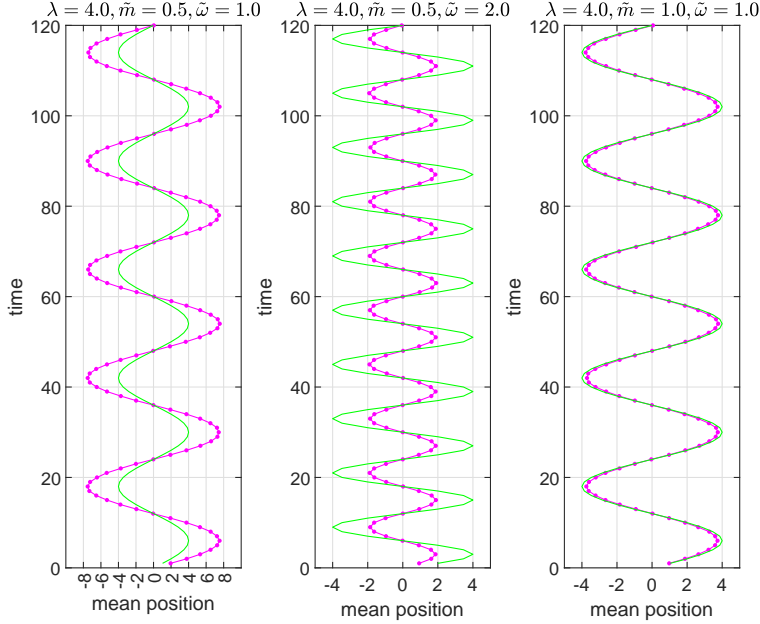}
    \caption{The solid (green) line shows the time-varying driving force. The dash-dotted line (magenta) shows the mean position at each point in the time lattice. We see that the time variation of the mean position is sinusoidal and resembles the equation \eqref{eq:fal}. The amplitudes of the sinusoids are equal to the maximum values of $\alpha_i$ from equation \eqref{eq:fal}. Equation \eqref{eq:fal} describes the position expectation value of a coherent state evolving in time. This shows that each simulated ground state is a coherent state with a position expectation value $\alpha_i$ at the time point $\tau_i$.}
    \label{fig:falpha}
\end{figure}

We have already seen the variation of $\alpha_i$ with $\lambda$ for a constant driving force in section \ref{section:mcmcfho}. In figure \ref{fig:falpha}, we have simulated the dependence of $\alpha_i$ on $\tilde{m}$ and $\tilde{\omega}$ for $\lambda = 4$. 
From equation \eqref{eq:fal}, the theoretical amplitudes of the position expectation values are $|\alpha_{i,max}|= 2 \lambda =8 $ for $\tilde m=0.5, \ \tilde \omega=1$, $|\alpha_{i,max}| = 0.25 \lambda$ =2 for $\tilde m=0.5, \ \tilde \omega=2$, and $|\alpha_{i,max}| = \lambda = 4$ for $\tilde{m} = 1, \ \tilde{\omega} = 1$. The simulated amplitudes of $\alpha_i(\tau_i)$ in figure \ref{fig:falpha} are very close to these the predicted values.

\section{Conclusion}
In the present study, we have simulated the ground state probability distributions of some forced harmonic oscillator potentials. Firstly, we chose a constant driving force. For each combination of the parameters $\tilde{m}, \tilde{\omega},$ and $\lambda$, the probability distribution of the waveform resembles a coherent state. The calculated position expectation values of the simulated states were found to be nearly identical to theoretical predictions. Subsequently, we considered a time-dependent sinusoidal forcing function and demonstrated that the ground state probability distribution is a coherent state that evolves with time. The position expectation value is also a sinusoidal function of time. 
It has the same frequency and initial phase as the driving force. We have also explored the dependence of this function on various combinations of $\tilde{m}, \tilde{\omega},$ and $\lambda$. The simulated values of the amplitudes closely match the predicted values. The methodology described here can be applied to any time-dependent forcing function. We can further attempt to simulate a forced harmonic oscillator where the mass and the natural frequency change with time. Such systems are used to formulate several physical phenomena, like the interaction of charged particles with time-varying electromagnetic fields\cite{charged}.  
Thus, evaluating the imaginary time path integral using MCMC methods is a powerful tool to visualize the ground state probability distributions of quantum systems. 

\begin{acknowledgments}
The authors thank professor Dr. Tanaya Bhattacharyya for her help and gratefully acknowledge the support of St. Xavier's College, Kolkata. 
\end{acknowledgments}

\bibliographystyle{plain}
\bibliography{fhogrub}



\end{document}